# Conceptual Entity-Relationship Model: Underneath the Simplicity and Staticity

**Sabah Al-Fedaghi** *

*Computer Engineering Department*
*Kuwait University*
*Kuwait*
**salfedaghi@yahoo.com, sabah.alfedaghi@ku.edu.kw**

*Abstract* – This paper deals with the issue of conceptual models' role in capturing semantics and aligning them to serve the remaining development phases of systems design. Specifically, the entity-relationship (ER) model is selected as an example of conceptual representation that serves this purpose in building relational database systems. It is claimed that ER diagrams provide a solid basis for subsequent technical implementation. The ER model appeal relies on its simplicity and its benefit in clarifying the requirements for databases. It is also claimed that the ER model has achieved a good equilibrium between expressive power on one hand and simplicity on the other. Nevertheless, some researchers have observed that this reduction of complexity is accompanied by oversimplification and overlooking dynamism. Accordingly, complaints have risen about the lack of direct compatibility between ER modeling and relational model. Substantial evidence exists showing that designers often provide incomplete, inaccurate, or inconsistent representations of domain features in the ER conceptual models they prepare. This paper is an attempt to explore what is beneath this static ER simplicity and its role as a base for subsequent technical implementation. In this undertaking, we use thinging machines (TMs), where modeling is constructed upon a single notion thimac (*thing/machine*). Thimac constituents are formed from the makeup of five actions, create, process, release, transfer, and receive—that inject dynamism alongside with structure. The ER's entities, attributes, and relationship are modeled as thimacs. Accordingly, in this paper, ER examples are remodeled in TM while identifying TM portions that correspond to ER components. The resulting TM model insets actions into entities, attributes and relationships. In this case, relationships are the products of creating linking thimacs plus the logic of constructing them. Based on such static/dynamic TM representation, the modeler can produce any level of simplification, including the original ER model. In conclusion, results indicated that the TM models facilitate multilevel simplicity and viable direct compatibility with the relational database model.

---------------------

\* Retired June 2021, seconded fall semester 2021/2022

## I. INTRODUCTION

In this paper, we concentrate on the issue of conceptual data models' role in capturing semantics and aligning them to serve the remaining development phases of systems design. Conceptual data modeling refers to the process of developing a representation of real-world (an application domain) data requirements. The focus is on the entity-relationship (ER) model and its variations, which are frequently used for the conceptual design of database applications in software engineering, business information systems, education, and research; in addition, many database design tools employ its concepts [1]. The ER model is very important for designing the logical structure of databases, and it helps in specifying data and relationships between the entities [1, 2, 3].

An ER model appeal relies on its simplicity [4] and its benefit in clarifying the requirements for the database [5]. Its diagrams create a "solid basis" for subsequent technical implementation [5]. According to Badia [4], "The ER model has achieved a good equilibrium between expressive power on one hand and simplicity and wide applicability on the other; any addition should be very well motivated, in the sense that it should be shown to be truly needed, with the benefits of adding it to the model clearly outweighing any drawbacks." Such proclaimed features of easiness, clarity, and expressiveness of the ER model deserve further analysis from the broad conceptual point of view.

On the other hand, some researchers have claimed that both the ER modeling and relational model are not directly compatible [6]. Substantial evidence exists to show that designers often provide incomplete, inaccurate, or inconsistent representations of domain features in the conceptual models they prepare [7]. There are situations where the ER model fails to capture semantics. Determining relationship sets may become difficult [8]. To incorporate additional semantics, modelers have proposed to extend the model with further features [4]; hence, exploring "enhancements and extensions to the ER model have become a legitimate and important area of research" [4].

This paper is an attempt to explore what lays beneath these ER model characteristics that have made it a conceptual base for subsequent technical implementation. In this undertaking, we use a diagrammatic modeling methodology based on thinging machines (TMs) [9].

### A. ER Model and Relational Database

The ER model has undergone a variety of changes and extensions over the years (e.g., [10]). The focus in this paper is on the basic ER model presented in Chen [11]. We assume an elementary knowledge of such a model and the relational databases. In this paper, generally, we will use mixed terminologies of ER and relational database.

The ER model describes data as entities, relationships, and attributes. These constructs are described in a rather intuitive way, and the distinction between entities and relationships is not always clear [12]. According to Chen [11], entity types are the ER counterpart of common nouns, and relationships are normally expressed by transitive verb phrases. Nouns converted from a verb correspond to relationships. Because verbs are typically the language proxy for events (including actions), the basic ER model is committed to the view that relationships are events [13].

A relational database is a representation of some aspect of the real world in terms of a collection of data elements. The relational database model has a modeling methodology independent of the details of the physical implementation [14]. A *relation* in a relational database (e.g., table) can be interpreted as a declaration; for example, STUDENT relation asserts that, in general, a student entity has a name, number, phone, and address. Each *tuple* (row) in the relation can be interpreted as a particular instance of an assertion.

### B. ER Difficulties

The ER model "is praised for having drawn the line between generality and expressive power at a good point, behind which we can expect diminishing returns" [4]. According to Kashyap [15], the ER approach helps one to arrive at a *true* or *complete picture* of the real world for which database is to be built, and it involves the identification and definition of entities of the concerned real world, entity grouping, and description, keeping in view the problem area context. ER modeling is described as "a key measure of success in the design of these models is the level that they accurately reflect the real world environment" [16].

Nevertheless, this reduction of complexity comes with oversimplification and the omission of dynamism. Accordingly, complains arise about the lack of direct compatibility between ER modeling and the relational model [6]. Substantial evidence exists showing that designers often provide incomplete, inaccurate, or inconsistent representations of domain features in the conceptual models they prepare [7].

Consider the notion of relations, which is one of conceptual modeling's most fundamental constructs [17]. According to West [28], real-world relationships do not automatically align with the lines in the ER graphs; hence, it would be confusing to use the word *relationship* for both.

According to Guarino and Guizzardi [13], the difference between entities and relationships is only a matter of a pragmatic modeling, or are there aspects of the intrinsic nature of "real-world" entities that would justify such distinction? Relationships are objects. Burton-Jones and Weber [7] mentioned that substantial evidence exists to show that designers often provide incomplete, inaccurate, or inconsistent representations of domain features in the conceptual models they prepare. Users often have difficulty understanding the meaning inherent in a conceptual model. Specifically, the grammatical construct of a relationship with attributes used in ER modeling produces ontologically unclear representations of a domain [7].

Additionally, the ER model and its extensions are generally lacking in constructs to model the dynamic nature of the real world [12]. According to Dey, Barron, and Storey [12], the ER model, at best, models a "snapshot" of the real world at any point in time; it does not contain specific constructs to model the dynamic aspects of the real world that reflect the state of affairs at a single time point; hence, it omits events. The incorporation of events as an addition to entities and relationships increases the semantic expressiveness of the resulting conceptual modeling language [12].

### C. Aims

This paper is an attempt to explore what is beneath this ER simplicity, clarity, and strength as a base for subsequent technical implementation. In this undertaking, we use a diagrammatic modeling methodology based on TMs. TM modeling is built upon the single notion: thimac (*thi*ng/*mac*hine). Thimac constituents are formed from the makeup of five actions: *create*, *process*, *release*, *transfer*, and *receive*. This is not a new idea; attempts to extend the ER model to involve dynamism have been proposed (e.g., [12, 18]) by introducing *events* as an additional construct. The TM model injects dynamism alongside a static structure using two-level ontology.

In analyzing the ER's entities, attributes and relationship, we propose to treat them uniformly as thimacs. Accordingly, actual ER models from the literature are remodeled in TM while identifying TM model portions that correspond to ER components. The resulting TM adds actions into entities and relationships. Relationships are the products of creating "linking thimacs" plus logic of constructing them. Based on such static/dynamic TM representation, we can produce any level of simplification, including the original ER model.

Fig. 1 shows a summary of work in this paper. On the left side, (a) we show a typical ER diagram and its mapping to the logical model of relational database. On the right side, we show the corresponding TM model where the entities, attributes, and relationships are thimacs.

### D. Sections

For the sake of a self-contained paper, the next section includes a brief review of the TM model's foundation of the theoretical development. Some materials in the section, such as the example, are a new contribution. Section 3 includes a discussion that demonstrates how to represent sets in TM modeling. Section 4 contains a case study of a sample ER model, taken from Captain [19], and its corresponding TM model. Section 5 contains a lager example of ER, taken from Elmasri and Navathe [2], and its modeling in TM. Section 6 comprises discussions of the notion of simplification in conceptual modeling.

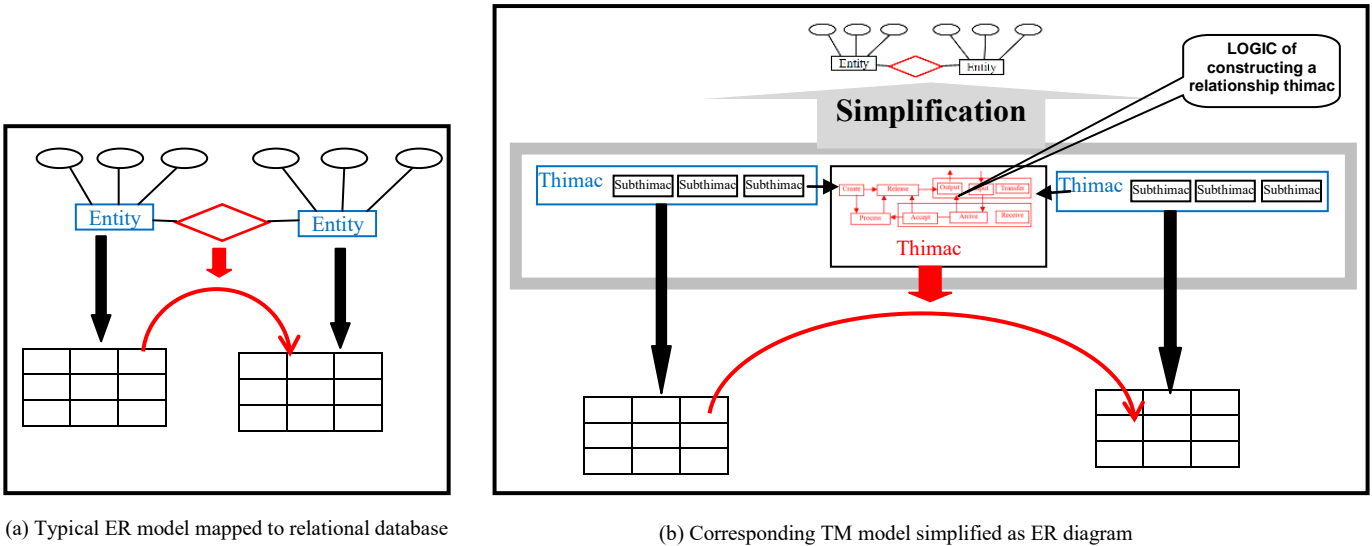

(a) Typical ER model mapped to relational database

(b) Corresponding TM model simplified as ER diagram

**Fig. 1 Overview of the work in this paper**

## II. TM MODELING

This section includes a summary of the TM model discussed in previous papers (especially recent publications, e.g., [10]). The TM model has evolved over the last 10 years, and model details can be found in many publications in different areas of applications such as physical security [20], unmanned aerial vehicles [21], information leakage [22], programming [23], customer relationship management [24], privacy [25], phone communication system [26], and documenting network systems [27]. In this paper, we apply TM modeling in set theory.

### A. General Outlook

The TM model provides us with an ontological representation of reality. It represents the entities and processes in the targeted domain utilizing one notion, which we call a *thimac* (*thing/machine*). Fig. 2 shows the structure of a thimac. In TM, the world is divided into thimacs in which different thimacs overlay or combine to form the texture of the whole as a grand thimac.

A generic thimac is a gathering up of elements into a unity or synthesis of actions: create, process, release, transfer, and receive. The thimac constituents are formed from the makeup of these actions. An action is a unit of actionality. A TM diagram is called a *region* at the static-modeling level. Fig. 3 shows a picture that outlines the two levels of the TM scheme.

The synthesis of actions is applied to *events* (at the dynamic level; see Fig. 3). The general idea of the thimac notion and two-level static/dynamic space and time along the lines of Hermann Minkowski's idea that "space by itself, and time by itself, are doomed to fade away into mere shadows, and only a kind of union of the two will preserve an independent reality" [28].

In TM, there is no ontological distinction between concrete and conceptual things (e.g., mathematical concepts), and all are *thimacs* (as a thing): what can be created, processed, released, transferred, and received things. Additionally, the thimac (as a machine) may create, process, release, transfer, and receive things. All so-called entities, properties, and relationships are thimacs or subthimacs.

### B. The Thimac

The thimac has the dual nature of a thing and a machine. It is a thing that is subjected to five actions and a machine that acts on things. A thimac's actions, shown in Fig. 2, are described as follows:

1) *Arrive*: A thing arrives at a thimac.

2) *Accept*: A thing enters a thimac. For simplification, the arriving things are assumed to be *accepted* (see Fig. 2); therefore, *arrive* and *accept* actions are combined into the *receive* action.

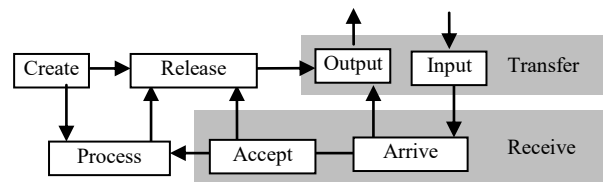

**Fig. 2. Thimac**

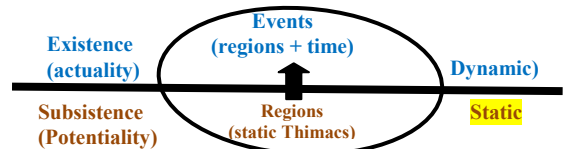

**Fig. 3 Two levels of TM modeling**

*3) Release:* A thing is ready for transfer outside the thimac.

*4) Process:* A thing is changed, handled, and examined, but no new thimac is generated.

*5) Transfer:* A thing crosses a thimac's boundary as input or output.

*6) Create:* A new thimac is registered as an ontological unit.

Additionally, the TM model includes *storage* (represented as a **cylinder** in the TM diagram) and *triggering* (denoted by a **dashed arrow**). Triggering transforms from one series of movements to another (e.g., electricity triggers heat generation).

### C. Two-Level Reality

The TM model has two modes of being—dynamic *existence* and static *subsistence*—that reflect a targeted portion of reality. A two-level reality is an old idea. According to traditional interpretations, there is the doctrine of degrees of reality in Plato's philosophy. The doctrine of degrees of reality says that forms exist, whereas particulars are half existent and half nonexistent [29]. Russell held that universals (e.g., "chairs" or "tigers") do not exist; they *subsist* and are nonetheless "something" [30]. According to Russell, subsistence is opposed to "existence" as being timeless. Thus, we must answer Parmenides's question, how can we talk about nonexistent objects? If we talk about them, it seems they must exist and that nonexistence things are subsisting things [31].

Fig. 3 defines the categorical structure of TM modeling. The two-level depiction is made to emphasize and illustrate the characteristics of each of the two levels; however, the two projected levels are superimposed over each other in TM modeling. Therefore, existence and subsistence are like a double-image impression (e.g., Rubin's vase), which is possible with a figure-ground perception. When we see an event, we simultaneously *perceive its region*. We seem to apprehend both at once. Such a phenomenon is facilitated by having the static region (essence) as a part of the event. Note that the so-called *objects* are a type of TM event. In TM, regions are entirely interior to events, but nothing that happens to an event can alter its region.

The TM static level is a world without time. Subsistence of regions resides in existence (of events). Time is immaterial to the definition of regions because they are hiding in events. This does not contradict the creation of new regions from previous regions.

### D. Examples

As an illustration of the TM two-level ontology, consider the issue of the nature of numbers. Aristotle said that numbers don't exist, but by "… the mode of their having being … the objects of mathematics are not substances to a greater degree than bodies nor prior in being to perceptible things" [32]. It is argued that a mathematical structure is an abstract, immutable entity existing outside of space and time (see Horne [32]).

According to Durante and Alves [33], a sample problem exists with application, as in 5+1=6, not yielding the same when applied to various situations (e.g., adding numbers of things not being physically the same).

In TM, numbers are thimacs represented by regions in the static level, as shown in Fig. 4, which represents 5+1=6. Fig. 4 includes four static thimacs. Thimacs 1 and 5 flow to the thimac that adds them to trigger the creation of the number 6. Note that *create* of 6 refers to the appearance of 6 in the context of the four thimacs models.

Fig. 5 shows how to represent "one cow" (e.g., this cow) as an event in the *existence* level. Number 1 and the thing "cow" are processed to form an *existing cow*. The thing "cow" and "1" *subsist* but do not *exist* by themselves and are involved in the dynamic process of creating "1 cow" or, say, "this is a cow." Such a notion of 'numbering things' as TM things can be generalized to 2, 3. 4, etc. All types of constraints are applied here (e.g., *'1 Cow'+'1 horse' = 2* cannot exist).

Note the TM modeling includes subsistence-based "truth." Here, a mathematical proposition describes the truth value of a subsistence reality (e.g., *5+1=6* is true). An "existence truth" example is *5cows+1cow=6 cows*.

The two events shown in Fig. 5 are a simplified representation of events. TM events are defined in terms of a static thimac plus time (e.g., Fig. 6).

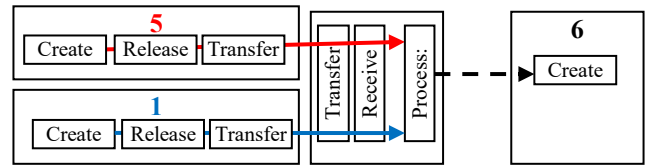

**Fig. 4    5+1=6 at the static level**

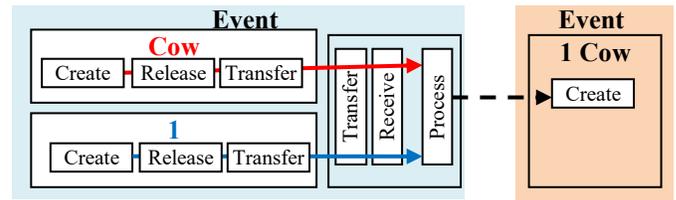

**Fig. 5 "One" cow exists in the TM model**

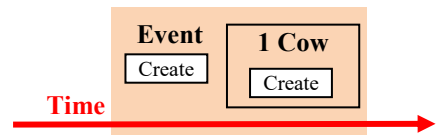

**Fig. 6 Definition of an event**

## III.    TM REPRESENTATION OF SETS

Set theory, which has become the standard foundation for mathematics, deals with well-determined collections, called *sets*, of objects that are called *members*, or *elements*, of the set.

## A. Sets in TM

In TM modeling, a set is a thimac (i.e., a thing and machine) [34]. As shown in Fig. 7, a set definition can be conceptualized as a thimac with three subthimacs:

- Member (singularity): one thing entering the set thimac to be integrated in the set extension.

- Extension (multiplicity): a pile of things, disregarding any order of the things that may be contained within it.

- A transformation between extension and member either to add a member to the set or taking out a member. For simplicity sake, this transformation involves mixed types, members, and extension (thick arrow) instead of retrieving members of the extension one by one.

A member can be received in the set (number 1), processed (2), and, if qualified, go to transformation (3). Additionally, the extension (4) is sent to the transformation. There, the member and extension are processed (5) to produce a new extension (6). Similarly, an extension (7) can be sent to the transformation to be processed to select a single member (8) that is sent to a member (9). The whole extension can be exported (10) [34].

## B. Dynamic Model

Fig. 8 shows the dynamic version of the set operations of Fig. 7.

The dynamic model includes the following events:

$E_1$: A set exits.

$E_2$: A member is received to be added to the set.

$E_3$: The input member is processed and found to be qualified to be a member in the set; thus, the new member flows to be added to the set.

$E_4$: The set is opened to be processed.

$E_5$: The new number is added to the set.

$E_6$: A new version of the set that includes the new member is created.

$E_7$: The set is opened.

$E_8$: The set is processed.

$E_9$: A member of the set is retrieved. Note the member is created in the sense that it does not appear in the system before its retrieval.

$E_{10}$: The whole set is released.

Fig. 9 shows the chronology of events of set operations.

Note that, typically, a database management system (DBMS) (a collection of programs that enables users to create and maintain databases and control) is a distinct component of the database; however, in TM database systems, the TMs integrate the static and dynamic aspects.

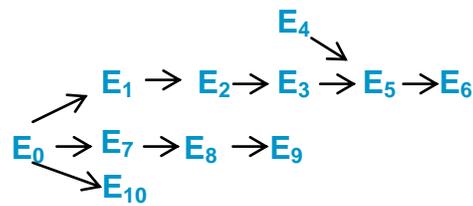

**Fig. 9 Chronology of events**

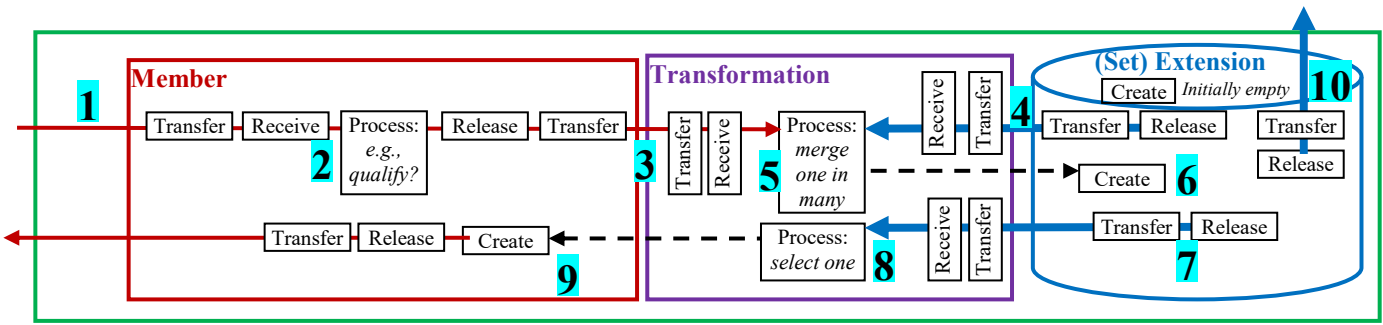

**Fig. 7 A set system definition**

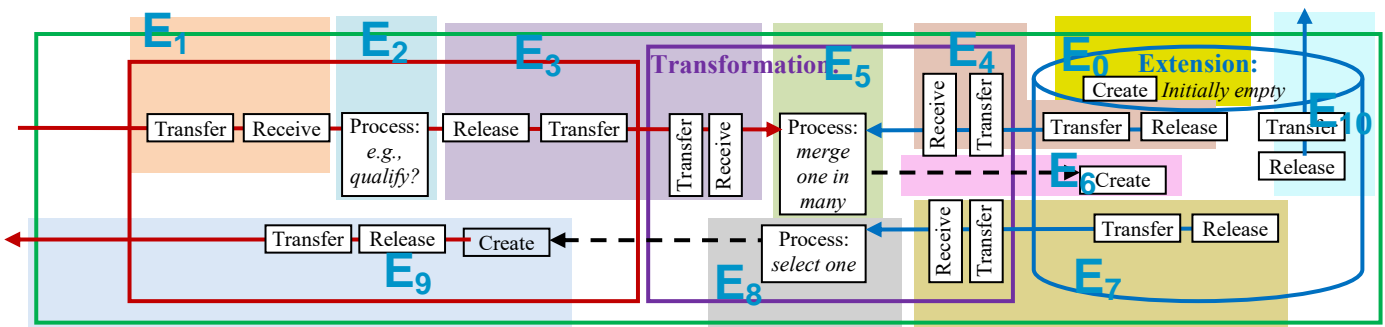

**Fig. 8 Input/output operations of a set**

## C. *Sets and Tables*

In TM modeling, relations (tables) in a relational database can be interpreted as thimacs realized from sets of tuples. In the rest of this paper, we will use a simplified representation of the definition of a set suitable for tables as sets of tuples. A set includes a collection of subthimacs representing members of the set, as shown in Fig. 10, which shows the set diagrammatic representation. The number of members of the set can be any number; in Fig. 10, the two members representation is a notational convenience. In Fig. 10, the *Create* action may be deleted to simplify the drawing, assuming that the rectangle is enough to indicate the creation of the set. For illustrative purposes, the rectangle of a set or subset may be drawn as a cylinder, as illustrated in Fig 11. Fig. 12 shows subsets of a set.

In mathematics, we create sets by placing all the elements inside curly brackets {}, separated by commas. Similarly, in TM, the set A is created in the TM potential level by inputting its members inside the thimac.

## IV. TM MODELING: ER CASE STUDY 1

This section includes a case study of a sample ER model and its corresponding TM model. Captain [19] provides an example of ER using the following scenario.

> Jim is a lecturer at a community college, and he wants an application to help him to keep track of the students he is advising. He wants to know which students register for which courses in which semester and what grade they got in those courses. He also wants the application to be password protected, with each user having a different login and password, and it must keep track of user logins/outs.

It is required to design a database that captures and stores all the relevant persistent data that will be used by the application. Fig. 13 shows one of two ER relationships of the result of the first step in the relational database design process. In contrast, in TM, we first develop the complete model, and not vice versa, as in the ER modeling where simple diagrams are used as a base for development. The TM model is produced as follows.

## A. *Static TM Model*

Fig. 14 shows the corresponding TM static model. In Fig. 14,
- The user enters his/her login ID (number 1) that flows to the system (2).
- Additionally, the users' file (as a set of tuples: 3) is processed (4) to extract a record (one by one), and an error message is sent if the end of the file is reached (5). The repeated retrieval process will be specified in the dynamic model later.
- The retrieved record is processed (6) to extract and save its ID within the record (7). Note that the *Create* action, here, indicates the appearance of the record or ID in the system after being embedded in the file or record, respectively. The password is also extracted and saved (the bottom left corner of the figure).

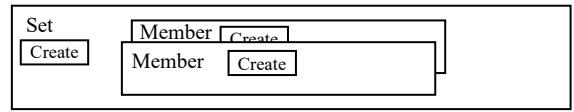

**Fig. 10 Diagrammatic representation of a set**

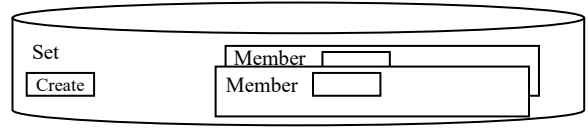

**Fig. 11 A set may be drawn as a cylinder**

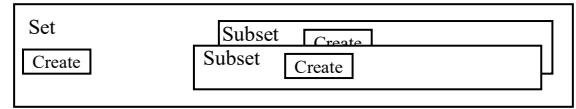

**Fig. 12 A set as a collection of subsets**

- The extracted ID (8) and input ID (9) are compared (10):
  (a) If the two IDs are not the same (11), then the next record is retrieved. Without a loss of generality, we assume here a sequential search process.
  (b) If the two IDs are the same (12), then the ID is saved for later use, and the system requests a password (13).
- The user inputs the password (14).
- The received password in processed (15) and is compared with the password of the extracted record (16; lower left corner of the figure).
  (a) If the two passwords are not the same (17), then an error message is sent.
  (b) If the two passwords are the same (18), then this triggers a user's session (19) and saves its starting time *LogON* (20).
- When the user session ends (21), then the record *LogEntry* is created (22), and the saved ID (23), starting time *LogOn* (24), and end time *LogOUT* are (25) inserted in the record *Logentry*.
- The *Logentry* record (26) and the *Logentries* file (27) are processed to create a new version of the file that contains the new record (28).

The static model of Fig. 14 can be simplified by assuming that the actions *release*, *transfer,* and *receive* can be eliminated based on the understanding that the direction of arrows is sufficient in representing the flow of things (See Fig. 15).

Further simplification can be achieved by eliminating the user, errors, and remaining actions such that it can be mapped to the ER diagram (See Fig. 16).

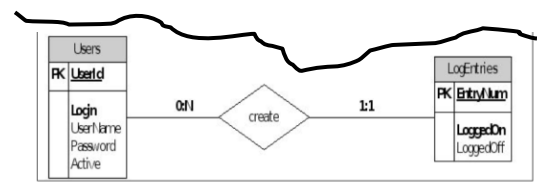

**Fig. 13 A relationship between two entities (Partial from Captain [19])**

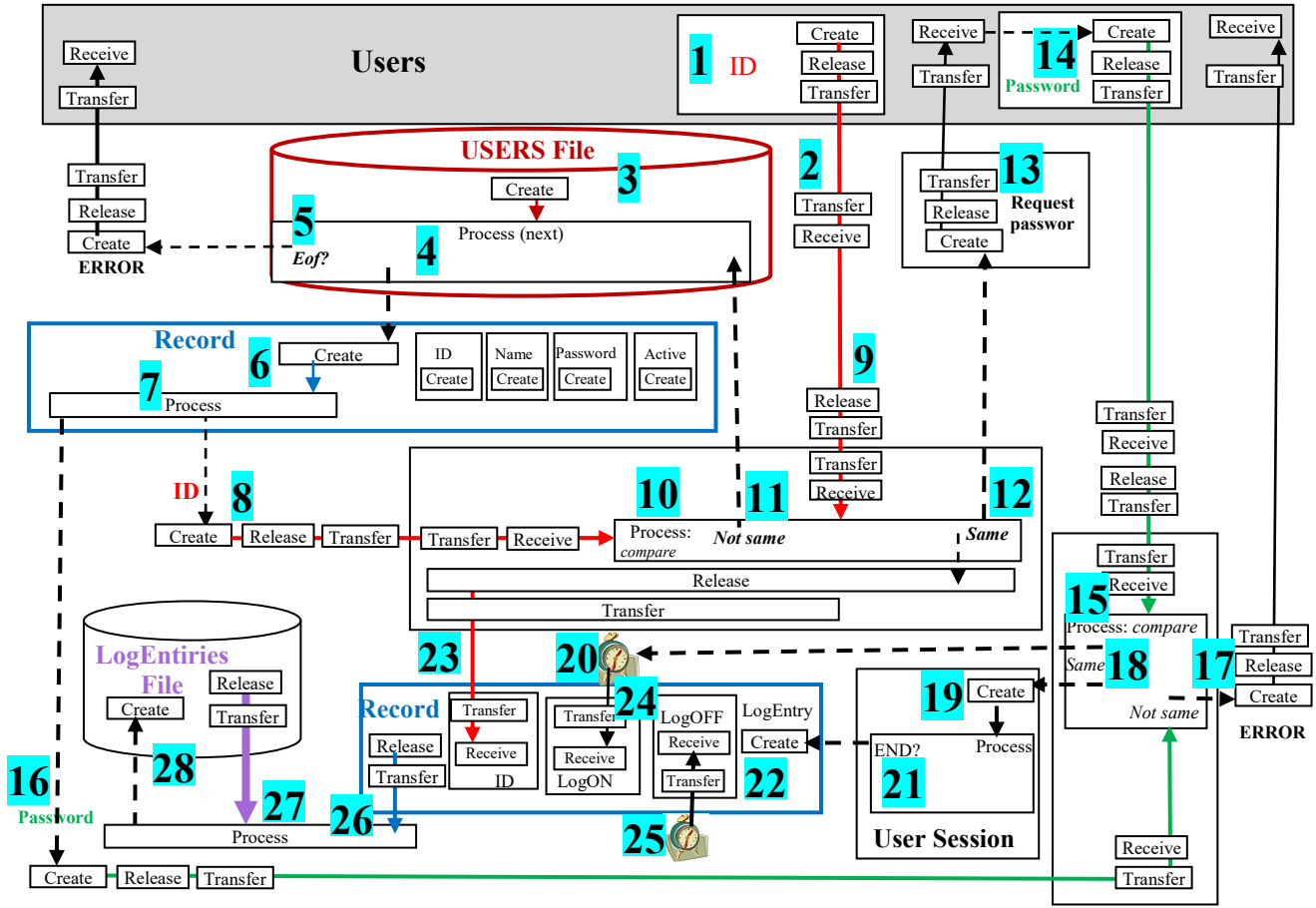

**Fig. 14 Static TM model**

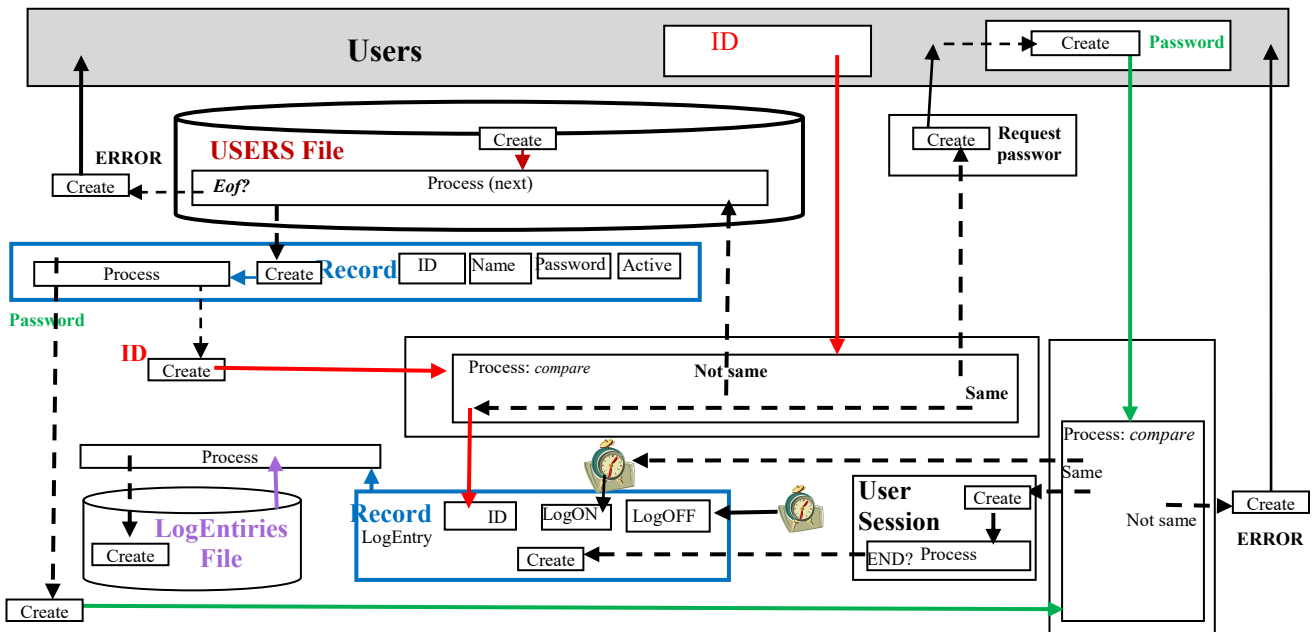

**Fig. 15 Simplified static TM model**

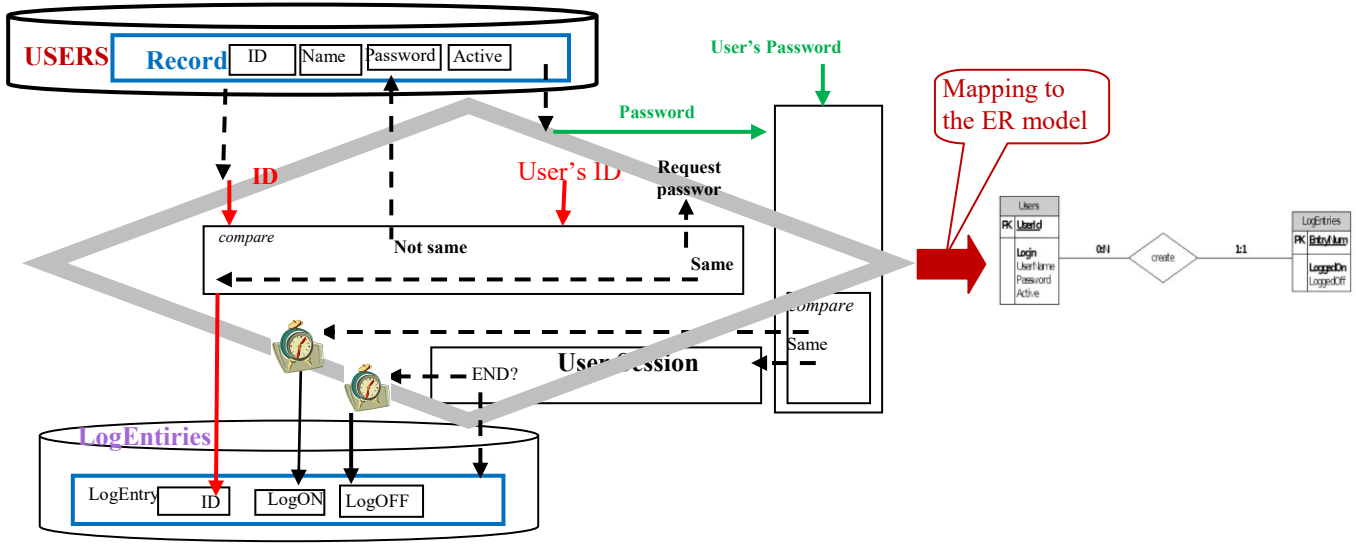

**Fig. 16 Simplification of the TM model to map it to the ER diagram**

Furthermore, to focus on the parts that relate *Users* and *LogEntiries*, we assume that the input ID is found in a record in *Users* and that the input password is a correct password. We call the involved operations in such a process the "logic" of relating *Users* to *LogEntry*, as shown in Fig. 16.

*B. Dynamic Model*

Fig. 17 shows the dynamic model where each event is represented by its region.

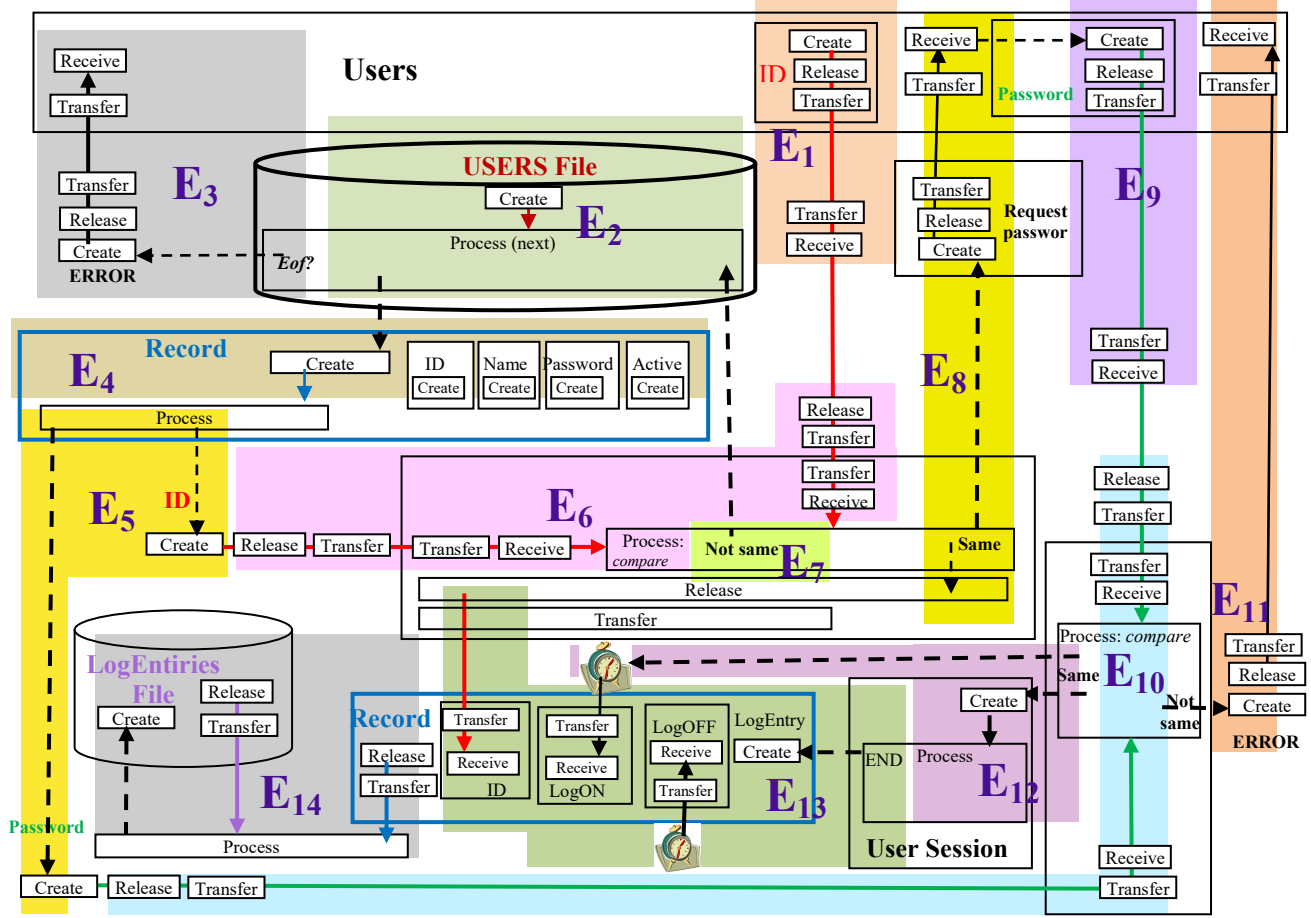

**Fig. 17 Dynamic TM model**

E₁: The user inputs his/her ID to the system.

E₂: The users file is processed.

E₃: Reaching the end of the file.

E₄: A record is extracted from the file.

E₅: The record is processed, and the ID and password are extracted and saved.

E₆: The extracted ID and input ID are processed.

E₇: The extracted ID and input ID are not the same.

E₈: The extracted ID and input ID are the same, and a request for a password is issued and sent to the user.

E₉: The user input a password.

E₁₀: The user's password and the stored password are compared.

E₁₁: The user's password and the stored password are different; hence, an error message is the output.

E₁₂: The user's password and the stored password are different; hence, a user's session starts, and its starting time are recorded.

E₁₃: The user's session ends; hence, a record *LogEntry* is created, and the ID, *LogON* time, and *LogOUT* time are inserted in the record.

E₁₄: The *LogEntry* is inserted in the *LogEntries* file.

Fig. 18 shows the chronology of events.

## V. TM MODELING: ER CASE STUDY 2

This section shows a larger example of ER and its modeling in TM. Fig. 19 shows an ER diagram taken from Elmasri and Navathe [2] with some omissions in order to reduce the size of the problem. Fig. 19 involves three entities and four relationships. The attributes are limited to two or three attributes. The ER schema diagram in the figure is for a company database. Some usual TM simplifications (e.g., rectangle without *Create*, using cylinders for files) will be applied in the corresponding TM model. Without loss of generality, some processes (e.g., sequential search) are assumed.

### A. Static Model

Fig. 20 shows the TM static model to be explained as follows: Color numbers will be used to point to elements in the diagram. Note that we will use different construction methods with different levels of description in building the logic of relationships in this example.

*The WORK_FOR relationship*:

- The model does not include the process of entering a record in a file. Thus, if EMPLOYEE (1) exists, relating it to DEPARTMENT (2) starts when a new record is inserted in EMPLOYEE and when requesting to assign it to an input department (3).

- The new record in EMPLOYEE is processed to extract the employee ID (4). Note that the area outside the entities and relationships is the realm of the system control. The ID flows to be part of the WORK_FOR record (5).

- On the other hand, the input department flows to the DEPARTMENT file to be processed (6) in order to be sure that the given DNum is one of the company's departments. The checking is not necessary in the context of relationships, and we can assume that the input DNum is always correct. However, we include it here to demonstrate aspects of the TM modeling. If the input is not one of the company's departments (e.g., large number), an error is outputted (7).

- If the department number is correct, it is extracted from the intended department (8) and sent to a process (9) where the employee ID and DNum form a record to be inserted in WORK_FOR.

- However, before doing that, the record is checked to determine whether it is already in WORK-FOR. This is accomplished by processing (ID, DNum) against WORK-FOR. This process (10) is simplified as involving WORK_FOR and the record (ID, DNum) in a loop (to be specified in the dynamic model).

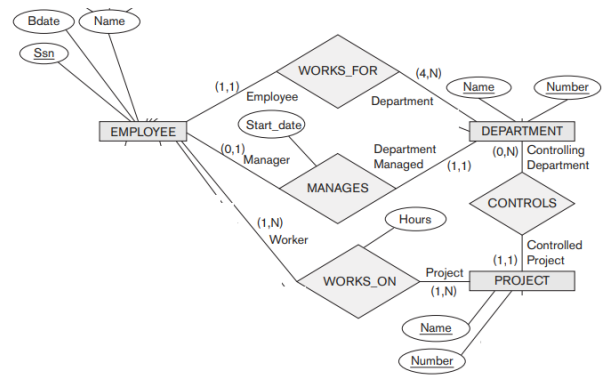

**Fig. 19 The ER schema diagram for the COMPANY database (Partial from Elmasri and Navathe [2])**

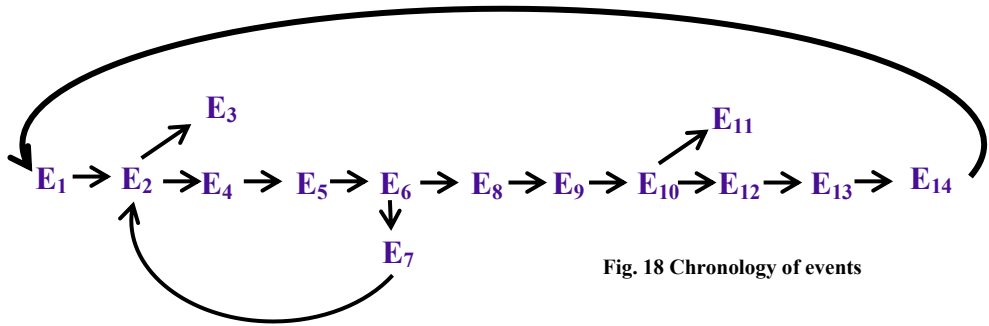

**Fig. 18 Chronology of events**

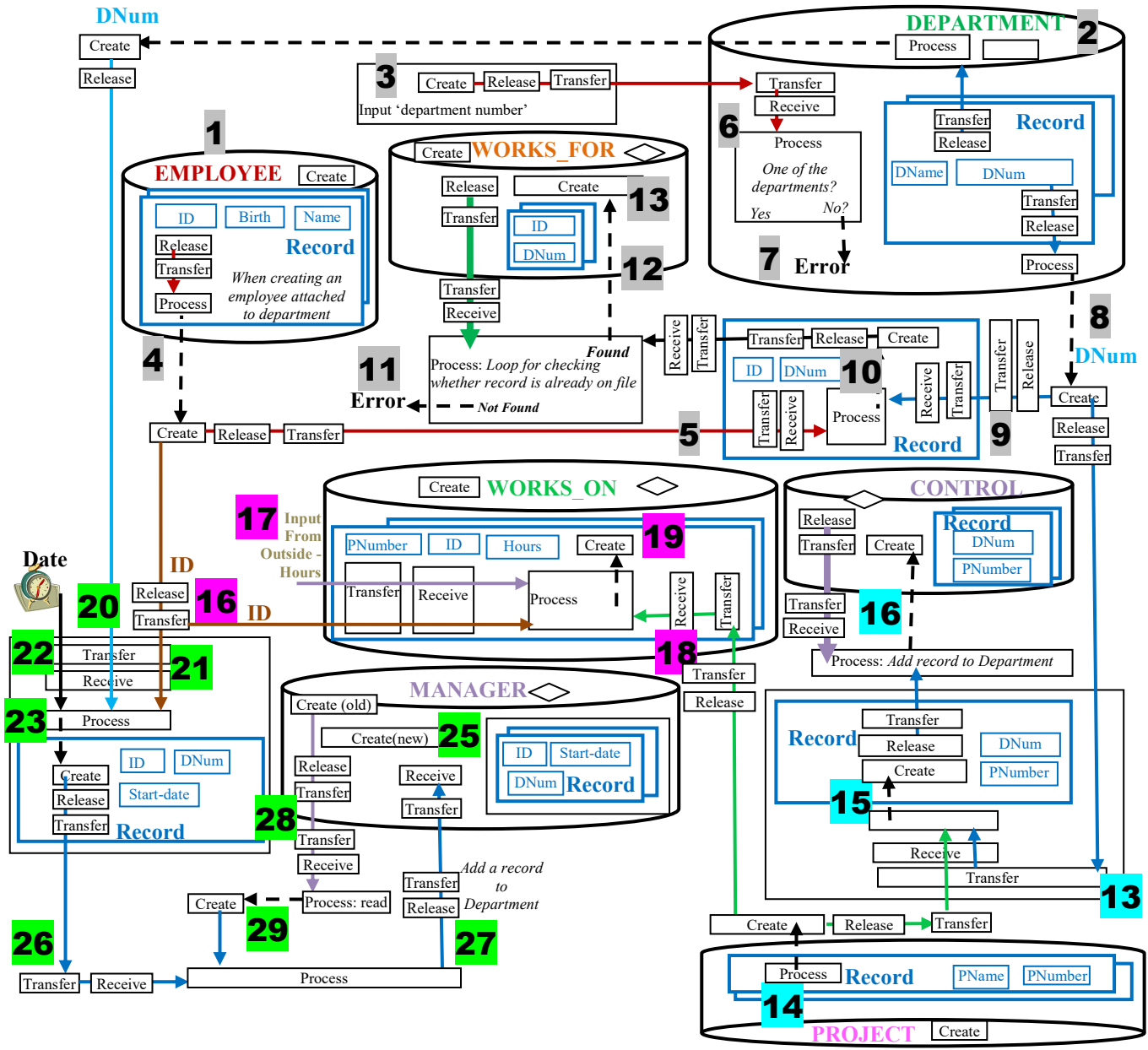

**Fig. 20 Static TM model**

However, the process could be specified in terms of a sequential search in WORK_FOR for the record. If the record is found in WORK_FOR, an error message is outputted (11).

- If the record is not WORK_FOR, then it is inserted in the file. Note that this means creating a new version of WORK_FOR (12).

*CONTROL Relationship*

This relationship is constructed simply by retrieving the DNum from DEPARTMENTS (13) and PNum from PROJECT (14), These values of attributes are processed to create a single record (15).

That record and the File CONTROLS are processed to create a new version of the file that includes the new record. The old file arrow is made thick to emphasize that the process of creating a new file has a heterogeneous input: a file and a record.

*WORK_ON relationship*

This relationship is constructed just like the preceding relationship, so we will not elaborate on the process of creating a record and then inserting it in the file. As it is modeled, it simply involves retrieving the ID (16), treating the Hours as the input (17), and utilizing PNum (18) to create a record (19).

*MANAGER relationship*

In constructing this relationship, the DNum, (20), (employee) ID (21), and the date (22) are processed (23) to create the record (DNum, ID, Start-date) (24). A new (empty) version MANAGER is open (25). The new version of MANAGER is filled with the new record (26 and 27), and then the records of the old version are retrieved (28 and 29) and inserted into the new version. The loop of performing this copying from the old version to the new version will be declared in the dynamic model. These sequential operations of inserting a new record in a new version are just for illustrative purposes. Any other method can be used.

Fig. 20 can be simplified, as we did in the previous section study case by eliminating the actions *release*, *transfer*, and *receive* under the assumption that the direction of the arrows indicates the flow of things. Consequently, Fig. 21 shows this first level of simplification.

**B. Dynamic Model**

Fig. 22 shows the dynamic TM model (using the static simplified version; Fig. 21) that includes the following. For simplicity sake, we assume there are no errors.

*WORK_FOR relationship events* (when a new employee):

$E_1$: EMPLOYEE as a set of employees exists (assuming it is initially empty).

$E_2$: Departments as a set of departments exist.

$E_3$: A new employee ID and an input DNum are processed.

$E_4$: A (ID, DNum) tuple is constructed.

$E_5$: WORKING_FOR exists (initially empty).

$E_6$: The (ID, DNum) tuple is not in WORKING_FOR.

$E_7$: The tuple (ID, DNum) is stored in WORKING_FOR.

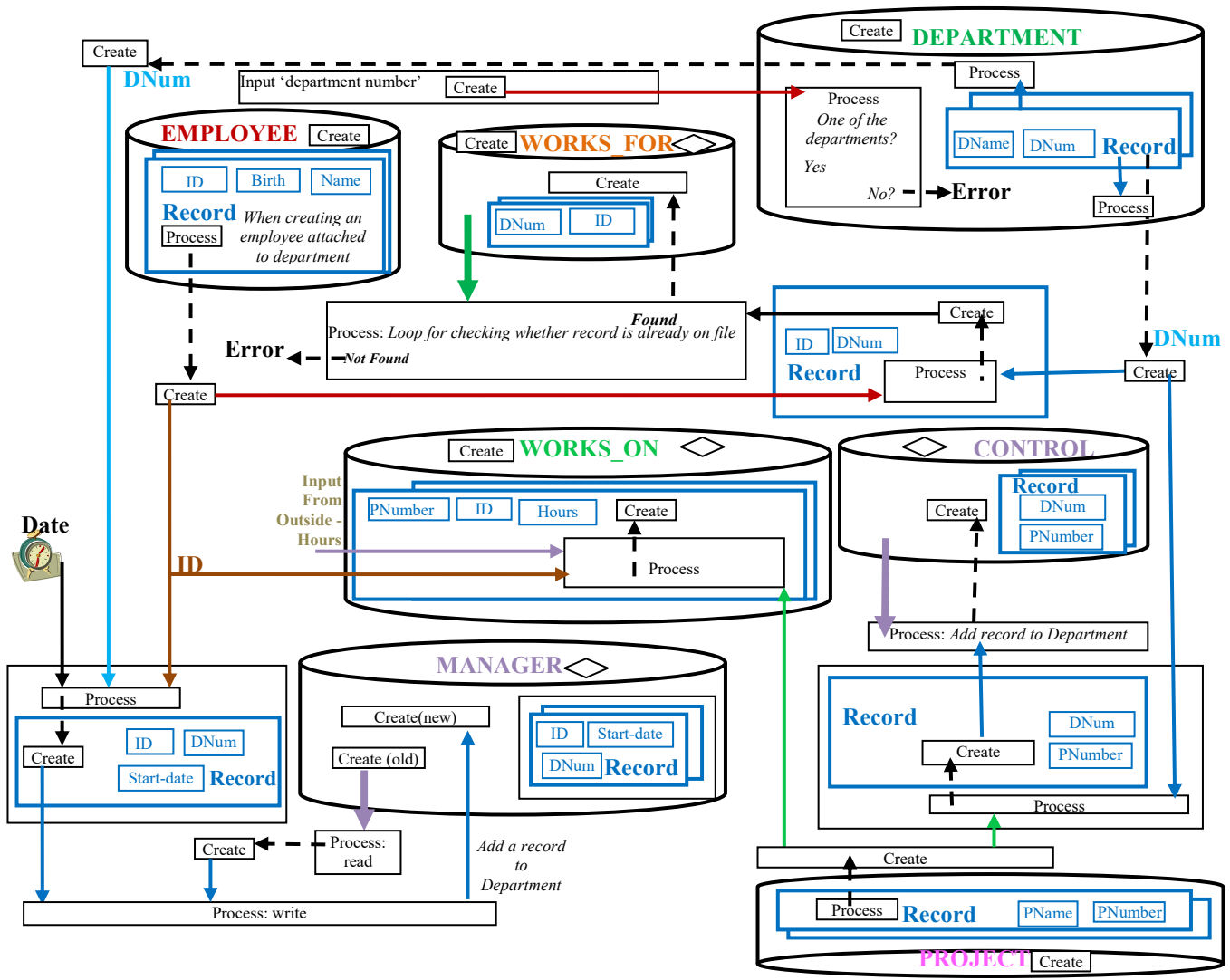

**Fig. 21 Static model simplification**

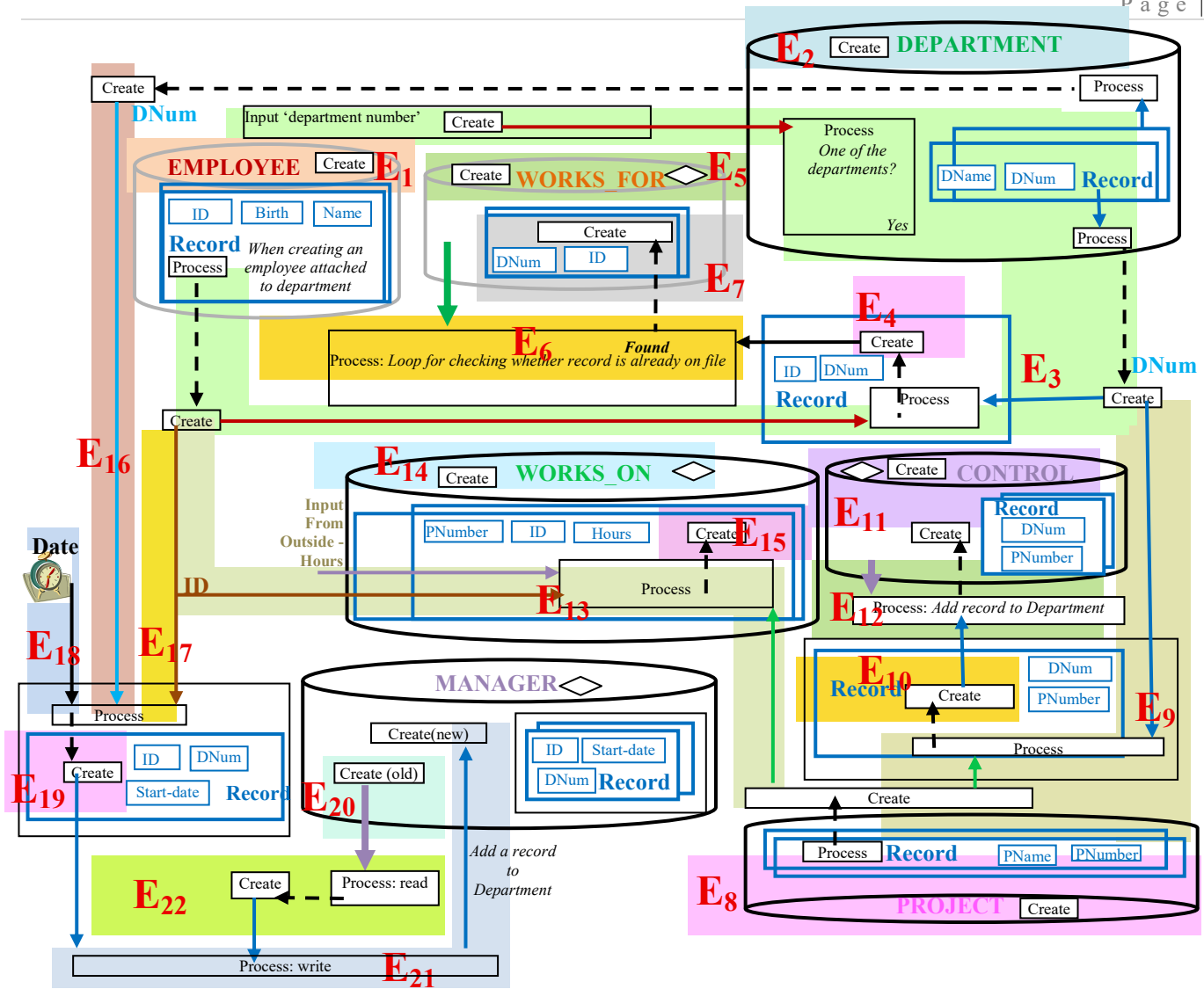

**Fig. 22 Dynamic TM model**

*CONTROL relationship*

$E_8$: PROJECT exists.

$E_9$: DNum is extracted from DEPARTMENT and PNum from PROJECT.

$E_{10}$: A (DNum, PNum) tuple is constructed.

$E_{11}$: PROJECT exists.

$E_{12}$: The A (DNum, PNum) tuple is stored in PROJECT.

*WORK_ON relationship*

$E_{13}$: ID and PNum are extracted From EMPLOYEE and PROJECT, respectively, and Hours is inputted.

$E_{14}$: A tuple (ID, PNum, Hours) is constructed.

$E_{15}$: The tuple (ID, PNum, Hours) is stored in WORK_ON.

*MANAGER relationship*

$E_{16}$: Dnum is extracted from DEPARTMENT.

$E_{17}$: ID is extracted from EMPLOYEE.

$E_{18}$: Start-date is received from the input.

$E_{19}$: The tuple (ID, DNum, Start-date) is constructed.

$E_{20}$: MANAGER exists.

$E_{21}$: Store the tuple (ID, DNum, Start-date) in a new MANAGER.

$E_{22}$: Copy data from the old Manager to the new MANAGER.

Fig. 23 shows the corresponding TM chronology of events. Finally, Fig. 24 shows the mapping of the TM model to an ER using the simplifications used in the previous case study.

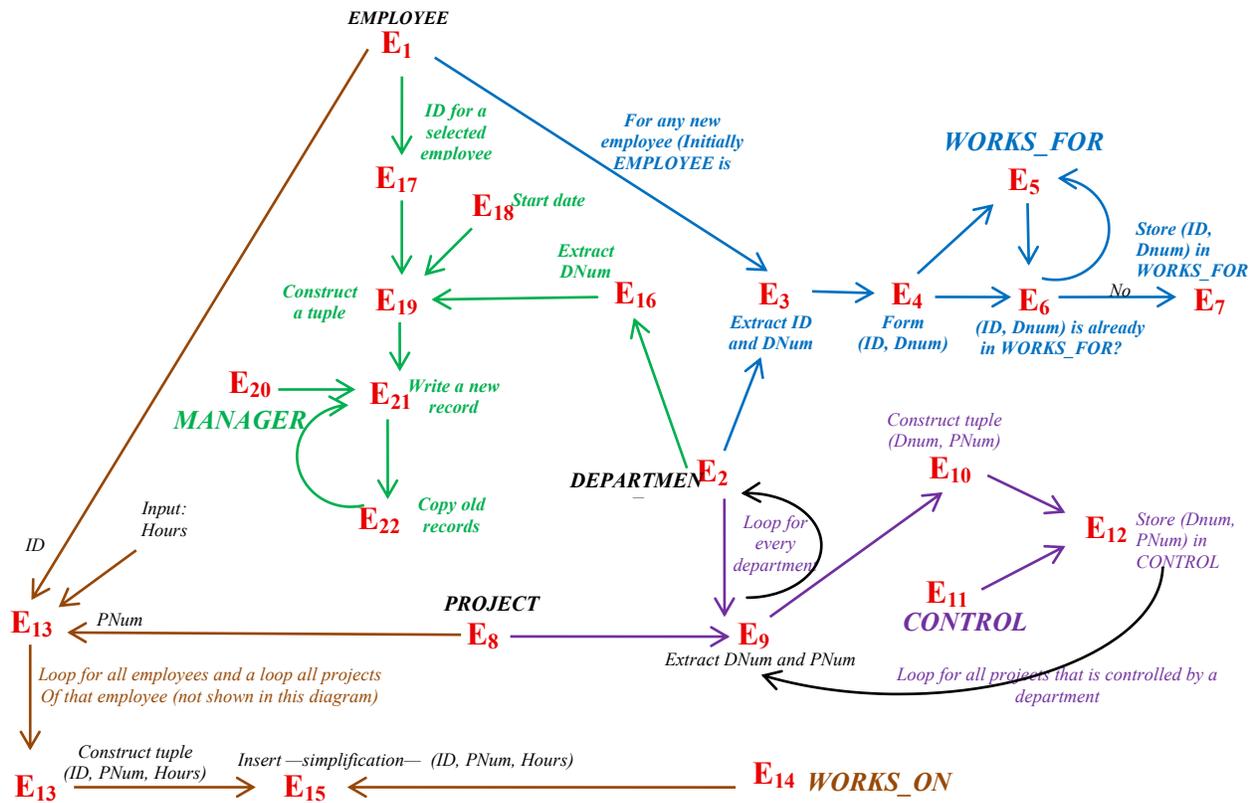

**Fig. 23 Chronology of events**

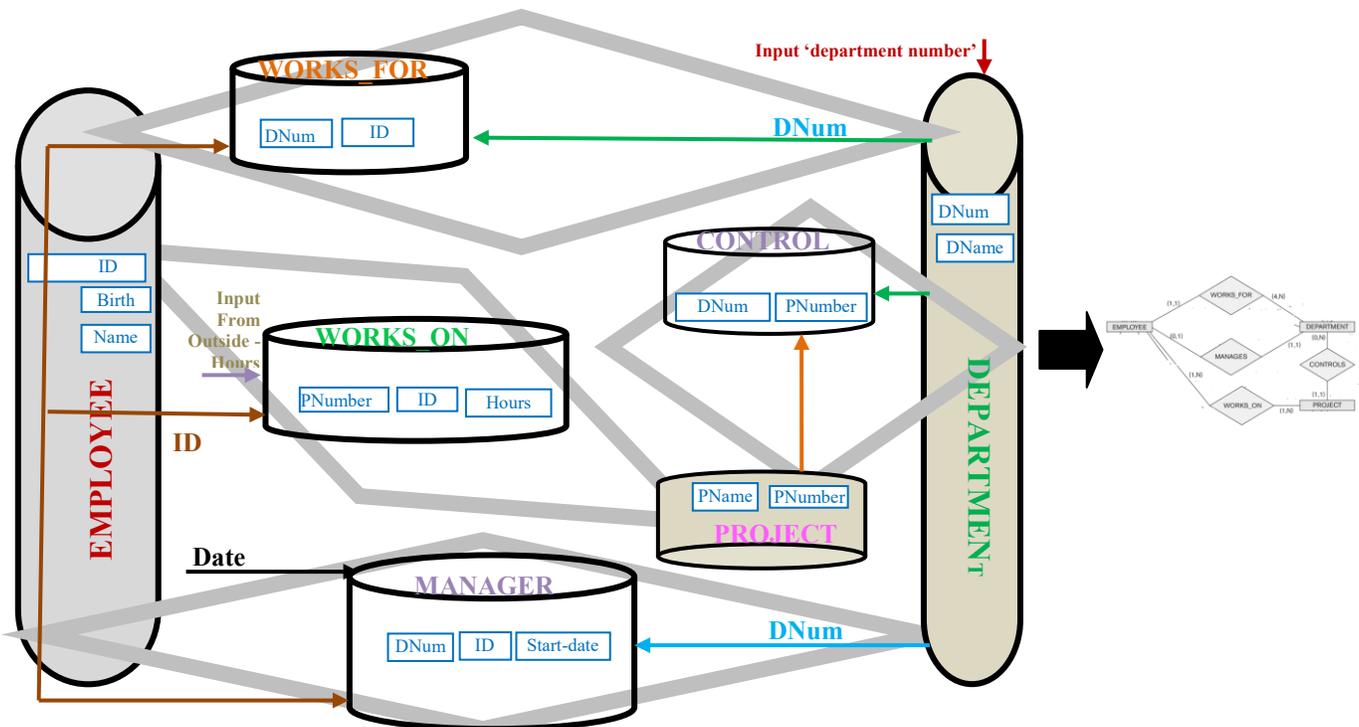

**Fig. 24 Further simplification that leads to the ER model**

## VI. DISCUSSIONS

Simplification is considered a fundamental part of modeling. Model simplification is instrumental in creating models that are useful—by focusing on system elements that matter—and feasible [35]. According to Lissack [36], simplification as a design choice only works up to a point. When simplification works, it can indeed be very effective. However, simplification does not always work, as it has a high risk of failure. Many times, simplification is *inappropriate*; it leads to outcomes that are poorly suited to the situation at hand. We then act based on the simplifications we have chosen, regardless of their *appropriateness* [36].

Simplicity may refer to syntactic simplicity, in measuring the number and conciseness of the theory's basic principles. On the other hand, ontological simplicity measures the number of kinds of entities postulated by the theory [37]. The main philosophical principle related to ontological simplicity is the well-known Occam's razor, which claims that entities should not be multiplied unnecessarily. The syntactical simplicity relates to the structure and organization [38]. The TM modeling has but one entity: the thimac with a structure based on the five generic actions that are connected through flows of things. This model can be simplified uniformly across its elements and structure using the following rules:

- Elimination of release, transfer, and receive under the assumption that the arrows indicate the direction of flows.
- Eliminating the action *create* under the assumption that the rectangles indicate the presence of the thimac in the model.
- Eliminating understood thimacs such as error messages in cases of unsatisfying conditions.

Such rules have been used in previous cases to move from the level of simplicity to a simpler one.

Additionally, the TM model uses the same modeling texture at the static and dynamic levels. This means that we recognize what aspects of the model are related to which aspects of reality: piece-by-piece correspondence or behavior-by-behavior correspondence [36].

Next, we demonstrate the simplicity of the TM modeling side by side with some ER diagrams.

*Example* (from Ask Question [39]): Consider the following scenario. Two completely different entities are independently related to the third entity in the same way:

- Student "BORROWS" BOOK (from the library).
- DEPARTMENT "BORROWS" BOOK (from the same library).

According to Ask Question [39], "How do we represent it in the ERD? or (Enhanced ER). If I define the "BORROWS" relationship twice, it would be awkward and clumsy in terms of appearance in the diagram, and increase the complexity of implementation as well."

We may have the following:

- A BORROWER entity (with the appropriate relationship), and have a STUDENT and DEPARTMENT subclass.
- To give borrowers relationships slightly different names. [39].

Fig, 25 shows the corresponding simplified TM model. It is simple, clear, and reflects the common sense meaning: the physical book moves to a student and to a department. Of course, the logic of establishing the two flows from BOOK is the next step of further modeling. At this level of simplicity, Fig. 26 shows the dynamic model and the chronology of events.

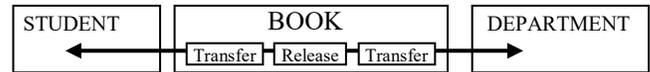

**Fig. 25 TM model**

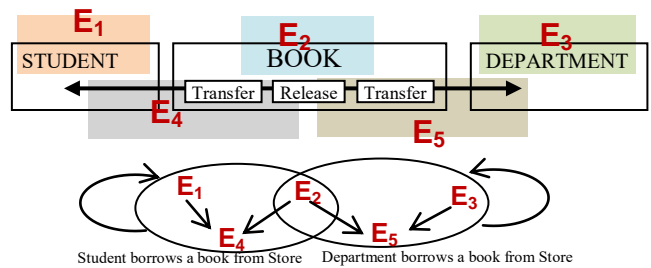

**Fig. 26 Dynamic model and chronology of events**

Suppose that we want to have a thimac that includes all borrowed books, regardless of whether the borrower is a student or department [39]. The STUDENT and BOOK thimacs can have the relationship thimac OR, as shown in Fig. 27. The keys of different thimacs depend on the semantics of interconnects. For example, borrowed book is identified with respect to its STUDENT and DEPARTMENT. OR can be, conceptually, a collection of two types of tuples (e.g., virtual table; see Fig. 27).

Fig. 28 and 29 show the events model and the chronology of events, respectively. Note that we can specify the type of relationship at the events level as shown in Fig. 30.

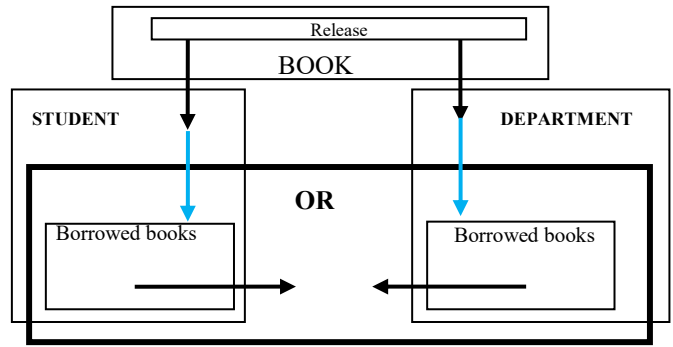

**Fig. 27 OR relationship that includes borrowed books by students or departments**

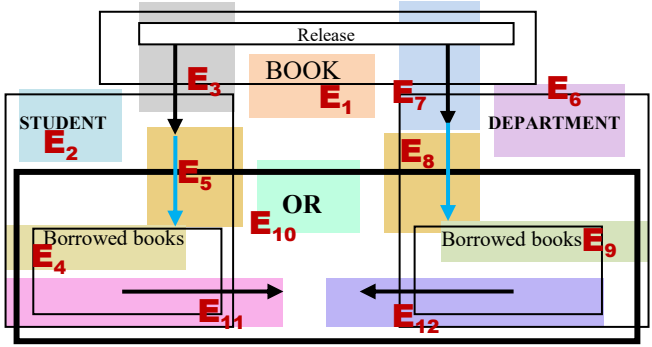

**Fig. 28 OR relationship that includes Borrowed books by students or departments**

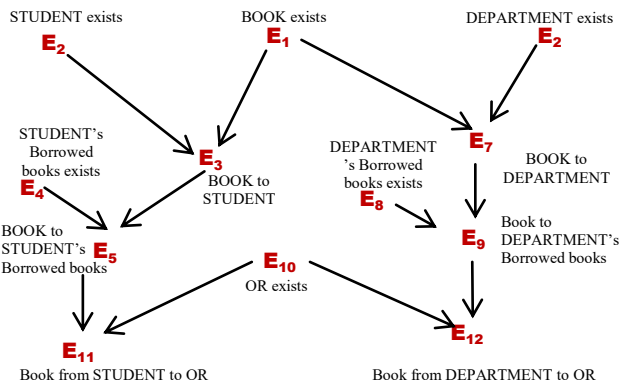

**Fig. 29 Dynamic model**

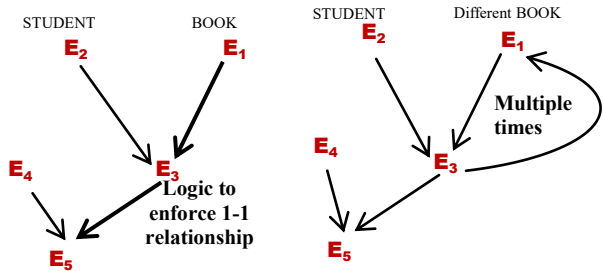

**Fig. 30 1-1 relationship (left) and 1-n relationship (right)**

*Example* (from Dullea, Song, and Lamprou [16]): Consider a situation where an association exists between the owner of a book, the title of the book, and the store where the book is purchased, and then a ternary relationship can be used to model this association (see Figs. 31 and 32). Note how the TM reflects the physical movement that raised the impression of an "association." For example, the association of a husband and wife in the TM model is depicted as the flow of husband and wife into marriage and not the flow of marriage into the husband and wife, as shown in Fig. 33.

According to Dullea, Song, and Lamprou [16], in Fig. 31, if we introduce an additional relationship that is independent of the OWNERSHIP relationship, such as the reader of the book, then the binary relationship READ between Person and Book in Fig. 31 is an explicit unrelated relationship. Owning and reading a book are two different concepts. Fig. 34 shows the corresponding TM model for this case.

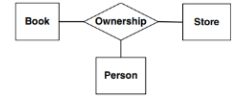

**Fig. 31 ER model**

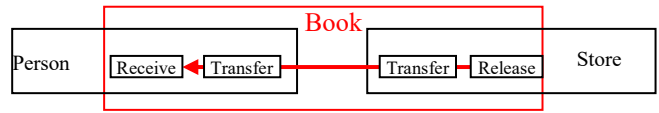

**Fig. 32 TM model**

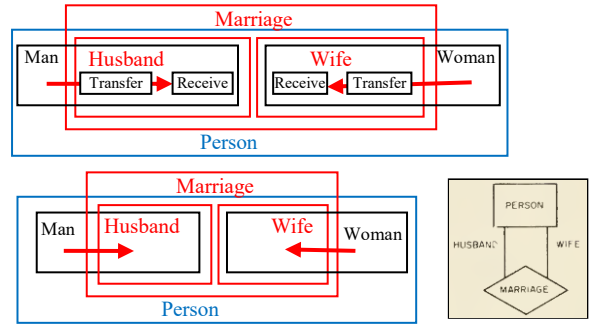

**Fig. 33 TM model of marriage (top), simplified version (bottom left) and ER diagram (bottom right) is from [11]**

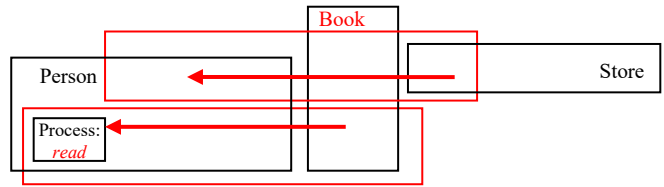

**Fig. 34 TM model after an additional relationship that is independent of the OWNERSHIP relationship**

## CONCLUSION

In this paper, we explored what is beneath the static ER simplicity regarding its role as a base for subsequent technical implementation. In this undertaking, we use TMs where modeling is constructed based upon a single notion: thimacs. We have demonstrated that the TM model offers degrees of freedom that provide the modelers the choice of the level of the system details suitable for naïve users and technical requirements. Based on the TM representation, the modeler can produce any level of simplification, including the original ER model.

Additionally, the TM model includes an extension of the basic static model to specify a system's dynamic behaviors. According to Lissack [36], "static descriptions are not true models" because "they do not provide any opportunity for us to simulate potential changes."

We proclaim that the TM model facilitates multilevel simplicity and viable direct compatibility with the relational database model. Further research will remodel more complex ER diagrams in TM to confirm this conclusion.